\documentclass[twocolumn,english]{IEEEtran}
\usepackage[T1]{fontenc}
\usepackage{babel}
\usepackage{prettyref}
\usepackage{float}
\usepackage{amsbsy}
\usepackage{amstext}
\usepackage{graphicx}
\usepackage[numbers,sort&compress]{natbib}
\usepackage[unicode=true,
 bookmarks=true,bookmarksnumbered=true,bookmarksopen=true,bookmarksopenlevel=1,
 breaklinks=false,pdfborder={0 0 0},pdfborderstyle={},backref=false,colorlinks=false]
 {hyperref}
\hypersetup{pdftitle={Your Title},
 pdfauthor={Your Name},
 pdfpagelayout=OneColumn, pdfnewwindow=true, pdfstartview=XYZ, plainpages=false}

\makeatletter

\providecommand{\tabularnewline}{\\}
\floatstyle{ruled}
\newfloat{algorithm}{tbp}{loa}
\providecommand{\algorithmname}{Algorithm}
\floatname{algorithm}{\protect\algorithmname}

\@ifundefined{date}{}{\date{}}
\ifCLASSOPTIONcompsoc
\usepackage[caption=false,font=normalsize,labelfont=sf,textfont=sf]{subfig}
\else
\usepackage[caption=false,font=footnotesize]{subfig}
\fi
\newrefformat{fig}{Figure\,\ref{#1}}

\setkeys{Gin}{width=\linewidth}
\usepackage{algorithmic}
\usepackage{xcolor}
\usepackage{undertilde}
\usepackage{cite}

\makeatother

\begin{document}
\title{Reduced Training Overhead for WLAN MU-MIMO Channel Feedback with Compressed
Sensing}
\author{Prasanna Sethuraman}
\maketitle
\begin{abstract}
The WLAN packet format has a short training field (STF) for synchronization
followed by a long training field (LTF) for channel estimation. To
enable MIMO channel estimation, the LTF is repeated as many times
as the number of spatial streams. For MU-MIMO, the CSI feedback in
the 802.11ac/ax requires the access point (AP) to send a null data
packet (NDP) where the HT/VHT/HE LTF is repeated as many times as
the number of transmit antennas $N_{t}$. With each LTF being $4\mu s$
long in case of VHT and $12\mu s$ to $16\mu s$ long in case of High
Efficiency WLAN (HEW), the length of NDP grows linearly with increasing
$N_{t}$. Furthermore, the station (STA) with $N_{r}$ receive antennas
needs to expend significant processing power to compute SVD per tone
for the $N_{r}\times N_{t}$ channel matrix for generating the feedback
bits, which again increases linearly with $N_{t}\cdot N_{r}$.

To reduce the training and feedback overhead, this paper proposes
a scheme based on Compressed Sensing that allows only a subset of
tones per LTF to be transmitted in NDP, which can be used by STA to
compute channel estimates that are then sent back without any further
processing. Since AP knows the measurement matrix, the full dimension
time domain channel estimates can be recovered by running the L1 minimization
algorithms (OMP, CoSAMP). AP can further process the time domain channel
estimates to generate the SVD precoding matrix.
\end{abstract}

\section{Introduction}

The WLAN technology since its commercialization at the turn of this
century has gained enormous popularity as WiFi and has become ubiquitous
- it has established itself in laptops, mobile phones and is seen
as the most promising way to connect IoT devices to the Internet.
The WLAN standard itself, driven by IEEE, has evolved from 802.11b
using direct sequence spread spectrum to 802.11a using OFDM \citep{OFDM-Reference,IEEE-802-11-Std}.
Multi-antenna transmission and reception (MIMO) was introduced in
802.11n while 802.11ac includes support for beamforming and multi
user MIMO (MU-MIMO). To enable beamforming, the beamformer needs to
know the channel that is seen at the beamformee. This requires the
beamformee to measure the channel and send it back to the beamformer.
Sending the channel matrix for each tone in feedback will result in
large overhead that increases linearly with MIMO dimension and number
of tones. In the cellular world, LTE, which is also based on OFDM,
handles this feedback problem by selecting a precoder from a predefined
quantized precoder set -- this minimizes the feedback to be sent,
but at the cost of being suboptimal. The feedback in 802.11ac computes
the SVD precoder matrix, which is then decomposed into Given's rotation
angles and sent back to the beamformer \citep{IEEE-802-11-Std}. The
beamformer can reconstruct the exact precoding matrix -- this scheme
thus provides the full gain that can be achieved by SVD precoding
and is therefore optimal, but at the cost of larger feedback. 

The newest version, 802.11ax High Efficiency WLAN (HEW), reduces the
sub-carrier spacing to efficiently support OFDMA and has support for
up to 8 transmit antennas. There are study groups at IEEE discussing
the next generation standard, called Extreme High Throughput (EHT),
where support for 16 or more antennas are being considered. With a
large number of antennas at the transmitter and receiver, sending
the full precoding matrix for each tone requires large number of bits
and the feedback overhead becomes prohibitively expensive. Furthermore,
computing SVD on the STA device also consumes power and is not desirable
for IoT type devices. Rather than low power devices losing out on
the beamforming gains, we could think about ways to reduce the feedback
overhead and if possible, move much of the processing from the STA
to the AP.

In WLAN, the overhead is not just in feedback, but also in the NDP
that must be sent to enable the beamformee to estimate the MIMO channel.
The number of LTFs that are sent depend on the number of transmit
antennas, and for large transmit antennas, significant air time is
occupied by sending multiple copies of LTF. 

In this paper, we propose two methods: the first of them is to find
a sparse representation for the channel jointly across both frequency
and spatial dimensions. Once we have such a sparse representation,
we can immediately reduce the number of feedback bits by sending only
the non-zero entries of the sparse channel. Even though this reduces
the feedback overhead, we still need to send full LTF pattern to estimate
the channel and transform it into sparse representation. The second
method that we propose uses the Compressed Sensing theory \citep{Compressed-Sensing,Candes-Intro-Compressive-Sampling}
so that we can reduce the feedback bits significantly without losing
information or optimality, and that the full precoder matrix can be
recovered at that AP with efficient algorithms. Using Compressed Sensing
allows us to have only small number of channel measurements enabling
us to reduce the size of the LTF.

The rest of the paper is organized as follows. We describe the system
model in section \ref{sec:System-Model} where the WLAN packet format,
LTF structure and the feedback process are explained. We also mention
the channel model and touch upon how the channel is sparse in time
domain. Section \ref{sec:Compressed Sensing} describes the Compressed
Sensing framework and the algorithms for $l_{1}$ minimization. Modeling
the channel estimation and feedback problem as a Compressed Sensing
problem is described in section \ref{sec:Proposed-Scheme}, where
we also describe how the LTF overhead can be reduced. We briefly discuss
the practical implementation aspects of this Compressed Sensing scheme
in \ref{sec:Implementation Complexity}. Conclusions are summarized
in sections \ref{sec:Summary}. 

\section{WLAN System\label{sec:System-Model}}

\begin{figure}[tbh]
\begin{centering}
\includegraphics{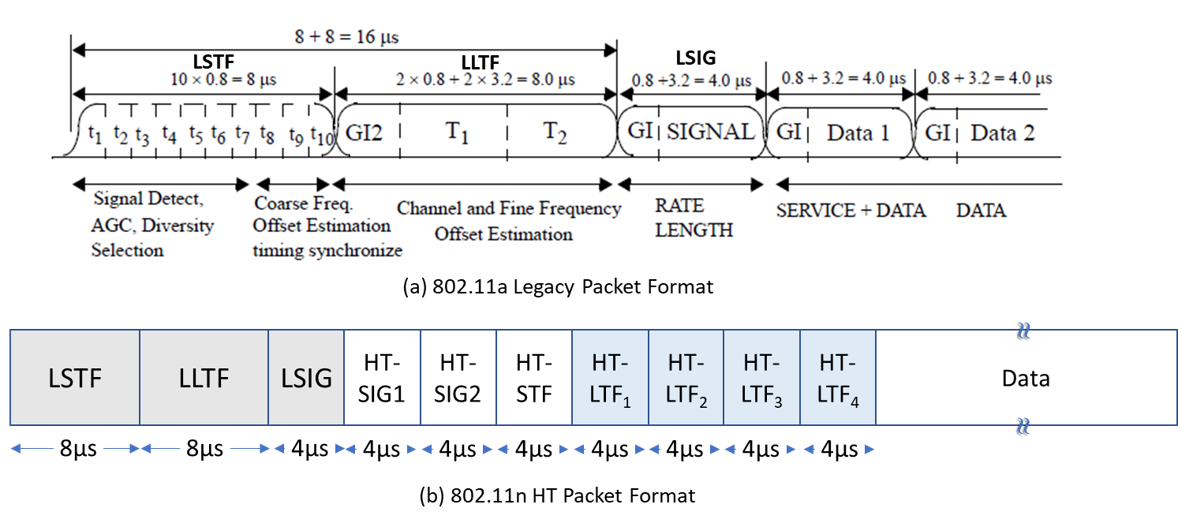}
\par\end{centering}
\caption{WLAN packet format\label{fig:WLAN-Packet-Format}}
\end{figure}

WLAN legacy packet structure is shown in \prettyref{fig:WLAN-Packet-Format}a
and has been extended in 802.11n ``high throughput'' PHY by adding
HT-LTFs for MIMO channel estimation. \prettyref{fig:WLAN-Packet-Format}b
shows the number of HT-LTFs in an 802.11n packet for four transmit
streams ($N_{ss}=4$). To enable the LTF sequence $L_{k}$ to be transmitted
from all available antennas, the IEEE 802.11 standard \citep{IEEE-802-11-Std}
specifies the $4\times4$ $\mathbf{P}$ matrix created by cyclic shifts
of the column vector $[1,1,1,-1]^{T}$. The transmit symbol $\textsc{x}_{k}(\ell,p)$
at tone $k$ antenna $\ell$ for $p$\textsuperscript{th} LTF and
is then generated by \prettyref{eq:TransmitLTF} where $[a(i,j)]$
denotes a matrix with $a(i,j)$ at row $i$ and column $j$. The received
signal for LTF $p$ tone $k$ at antenna $m$ can then be written
as in \prettyref{eq:ReceiveLTF}. The $N_{r}\times N_{t}$ MIMO channel
matrix $\mathbf{H}_{k}$ for tone $k$ can then be extracted with
\prettyref{eq:LTFChanEst}. For $N_{t}>4$, the standard introduces
$6\times6$ and $8\times8$ $\mathbf{P}$ matrices to support up to
$6$ and $8$ transmit antennas respectively. If this approach is
extended up to $N_{t}=16$ transmit antennas over 802.11ax HEW, we
would need additional $\mathbf{P}$ matrices and the duration of LTFs
alone would be as much as $16\times12\mu s=192\mu s$.
\begin{eqnarray}
[\textsc{x}_{k}(\ell,p)] & = & \mathbf{P}\cdot L_{k}\label{eq:TransmitLTF}\\{}
[\textsc{y}_{k}(m,p)] & = & \mathbf{H}_{k}\cdot\mathbf{P}\cdot L_{k}+[\eta_{k}(m)]\label{eq:ReceiveLTF}\\
\mathbf{\hat{H}}_{k} & = & [\textsc{y}_{k}(m,p)]\cdot L_{k}\cdot\mathbf{P}^{T}\label{eq:LTFChanEst}
\end{eqnarray}

\begin{figure}[tbh]
\begin{centering}
\includegraphics{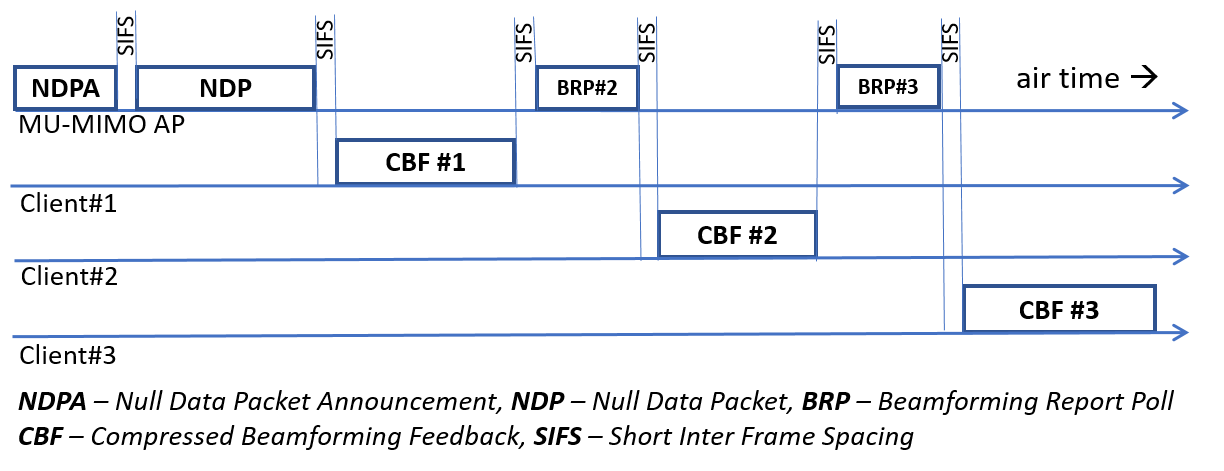}
\par\end{centering}
\caption{MU-MIMO Channel Sounding Procedure\label{fig:MU-MIMO-NDP-Handshake}}
\end{figure}

For the MIMO channel to be estimated, the beamformer sends a Null
Data Packet (NDP) which has the same format as \prettyref{fig:WLAN-Packet-Format}b
but without any data portion. \prettyref{fig:MU-MIMO-NDP-Handshake}
shows the channel sounding procedure for MU-MIMO feedback. The number
of HT/VHT LTFs transmitted in the NDP is equal to the maximum spatial
stream that can be supported by the beamformer, which usually is equal
to its number of transmit antennas $N_{t}$. The beamformer, on reception
of NDP, estimates the channel matrix $\mathbf{H}_{k}$ for all tone
$k$, computes the SVD $\mathbf{H}_{k}=\mathbf{U}_{k}\mathbf{\Sigma}\mathbf{V}_{k}^{H}$,
decomposes $\mathbf{V}_{k}^{H}$ into a series of Givens' rotation
angles, and sends back the angles as feedback. Table \ref{tab:Feedback-bits-per-tone}
shows the number of bits per tone for the quantized angles for different
transmit and receive antennas. For 80 MHz VHT, we need to send feedback
for $234$ tones and the total feedback bits then becomes $234\times864=202,176$
for a $N_{t}=16,\:N_{r}=4$ system.

\begin{table}[tbh]
\caption{Feedback bits per tone for quantized angles\label{tab:Feedback-bits-per-tone}}

\centering{}%
\begin{tabular}{|c|c|c|c|c|c|}
\hline 
Type & 2T2R & 4T2R & 8T2R & 16T2R & 16T4R\tabularnewline
\hline 
\hline 
Single User (SU) & 10 & 50 & 130 & 290 & 540\tabularnewline
\hline 
Multi User (MU) & 16 & 80 & 208 & 464 & 864\tabularnewline
\hline 
\end{tabular}
\end{table}
\begin{figure}[tbh]
\begin{centering}
\includegraphics{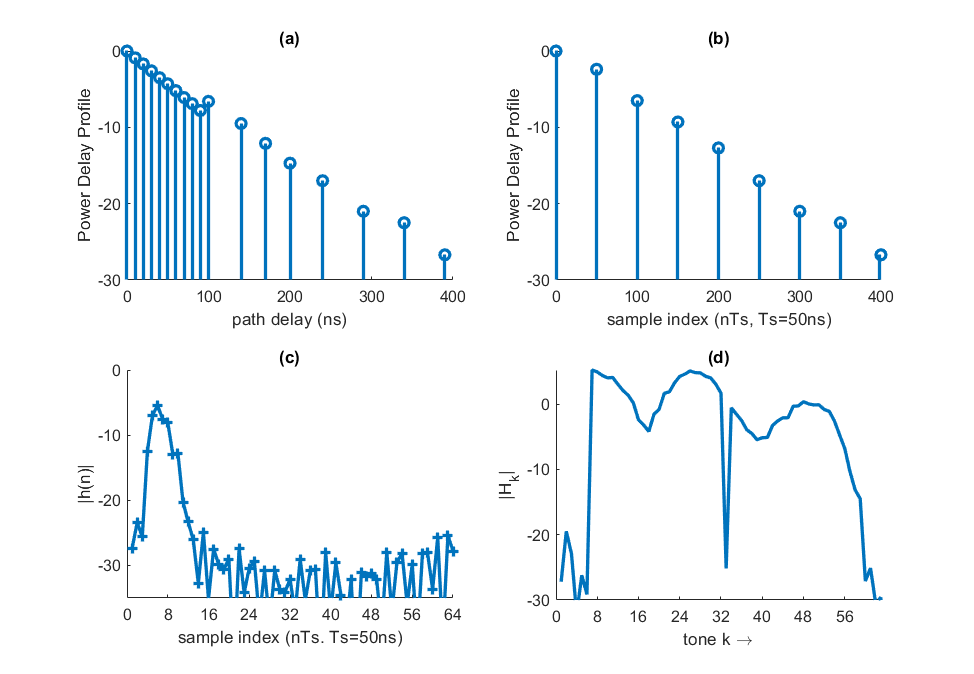}
\par\end{centering}
\caption{Channel model (a), (b) and measured time and frequency domain channels
(c), (d)\label{fig:ChannelPlot}}
\end{figure}

The $N_{r}\times N_{t}$ channel matrix $\mathbf{H}_{k}$ contains
the frequency domain channel $H_{k}(\ell,m)$ for each TX-RX antenna
pair $(\ell,m)$ which we can collect in a vector $\mathbf{H}_{\ell m}=[\begin{array}{cccc}
H_{1}(\ell,m) & H_{2}(\ell,m) & \cdots & H_{N_{DFT}}(\ell,m)\end{array}]$. The IFFT of this vector $\mathbf{H}_{\ell m}$ then gives us the
time domain channel $\mathbf{h}_{\ell m}$ between the TX antenna
$\ell$ and the RX antenna $m$. We can model this time domain channel
as a tapped delay line with time varying coefficients that follows
a Power Delay Profile (PDP). An example PDP of the ``model-D'' channel
defined in \citep{TGnChannelModel} is shown in \prettyref{fig:ChannelPlot}a.
Summing all the channel taps within one sample time $T_{s}=50ns$
for 20 MHz results in the PDP shown in \prettyref{fig:ChannelPlot}b.
One instance of the measured time domain and frequency domain channels
are shown in \prettyref{fig:ChannelPlot}c and \prettyref{fig:ChannelPlot}d
respectively. We see that the time domain channel is quite sparse,
since the channel delay spread is expected to be less than the cyclic
prefix duration. Therefore, instead of sending feedback on $52$ tones
for the 20 MHz system, we can only send the non-zero taps in the time
domain channel, which is less than one fourth of the number of tones.

\section{Compressed Sensing\label{sec:Compressed Sensing}}

Given a $N\times1$ signal vector $\textsc{\textbf{x}}$ is $\kappa$-sparse,
meaning it has only $\kappa\ll N$ non-zero elements, then the compressed
sensing framework allows recovery of $\textbf{x}$ from just $2\kappa$
measurements $\boldsymbol{y}$ obtained with a $2\kappa\times N$
measurement matrix $\boldsymbol{\Phi}$, provided $\boldsymbol{\Phi}$satisfies
the Restricted Isometry Property (RIP) -- that is, there exists $\delta_{\kappa}\in(0,1)$
such that $(1-\delta_{\kappa})\leq\|\boldsymbol{\Phi}\boldsymbol{\textsc{x}}\|_{2}/\|\boldsymbol{\textsc{x}}\|_{2}\leq(1+\delta_{\kappa})$
for all $\kappa$-sparse vectors $\boldsymbol{\textsc{x}}$. RIP intuitively
means that if distances are well preserved in the linear transformation
with $\delta_{2\kappa}<1$, then there are no two $\kappa$-sparse
vectors that will result in the same measurement vector $\boldsymbol{y}$.
Furthermore, the recovery of $\boldsymbol{\textsc{x}}$ from $\boldsymbol{y}$
can be achieved by $l_{1}$ minimization: $\hat{\boldsymbol{\textsc{x}}}=\min\Vert\boldsymbol{\textsc{x}}\Vert_{l_{1}}:\boldsymbol{\Phi}\boldsymbol{\textsc{x}}=\boldsymbol{y}$.
There are greedy algorithms -- Orthogonal Matching Pursuit (OMP)
and Compressed Sampling Matching Pursuit (CoSaMP) for example --
that can solve this $l_{1}$ minimization problem. For this work,
we use CoSaMP \citep{CoSaMP} for signal reconstruction. 
\begin{algorithm}[tbh]
\caption{CoSaMP\label{alg:CoSaMP}}

\begin{algorithmic}[1]
\STATE Inputs: $\mathbf{\Phi}$, $\mathbf{y}$, $\kappa$, $\tau$, $I_{max}$
\STATE $\mathbf{r} = \mathbf{y}$, $i = 0$, $T=[\;]$, $\hat{\mathbf{x}}=\mathbf{0}_{N\times 1}$
\WHILE{$i\leq I_{max}$ \AND $\Vert \mathbf{r}\Vert_2/\Vert \mathbf{y}\Vert_2 > \tau$}
\STATE $\mathbf{u} = \mathbf{\Phi}^H\mathbf{r}$ \\\COMMENT{\textcolor{blue}{$\mathbf{\Phi}$ is $N_\kappa\times N$, $\mathbf{r}$ is $N_\kappa\times 1$, $\mathbf{u}$ is $N \times 1$}}
\STATE $\Omega = support(\mathbf{u},2\kappa)$ \\\COMMENT{\textcolor{blue}{$support(\mathbf{u},2\kappa)$ has locations of $2\kappa$ largest elements of $\mathbf{u}$}}
\STATE $T = T \cup \Omega$ 
\STATE $\mathbf{b}=\mathbf{\Phi}_{(T)}^\dagger\mathbf{y}$ \label{CoSaMP-pseudo-inv} \\\COMMENT{\textcolor{blue}{$\mathbf{\Phi}_{(T)}$ is the submatrix of $\mathbf{\Phi}$ with only entries selected by $T$ and $\dagger$ denotes the pseudo inverse}}
\STATE $T = T_{(support(\mathbf{b},\kappa))}$ \\\COMMENT{\textcolor{blue}{Pruning the least-square solution $\mathbf{b}$. Get the indices corresponding to $\kappa$ largest elements of $\mathbf{b}$}}
\STATE $\hat{\mathbf{x}}_{(T)}=\mathbf{b}_{(T)}$
\STATE $\mathbf{r} = \mathbf{y} - \mathbf{\Phi}_{(T)}\mathbf{b}_{(T)}$
\STATE $i=i+1$
\ENDWHILE
\end{algorithmic}
\end{algorithm}

CoSaMP, similar to the other matching pursuit algorithms, tries to
find which basis vectors of $\boldsymbol{\Phi}$ that have the largest
dot product with the measured signal $\boldsymbol{y}$. In other words,
$\boldsymbol{y}$ has maximum contribution from these basis vectors.
So these basis vectors give an estimate of which elements of the signal
vector $\boldsymbol{\textsc{x}}$ are non-zero. We then extract these
basis vectors to find the least square estimate $\hat{\boldsymbol{\textsc{x}}}$
of $\boldsymbol{\textsc{x}}$. Check how close this estimate is to
$\boldsymbol{y}$ by computing the residue $\boldsymbol{r}=\boldsymbol{\Phi}\hat{\boldsymbol{\textsc{x}}}$.
If the residue $\boldsymbol{r}$ is not close enough to $\boldsymbol{y}$,
repeat the steps with $\boldsymbol{r}$ until convergence. 

It is worthwhile to note here that while Compressed Sensing requires
at least $2\kappa$ measurements to recover a $\kappa$-sparse signal,
we will be dealing with complex signals and therefore $\kappa$-sparse
means there are $2\kappa$ non-zero entries in the signal vector.
We therefore need at least $4\kappa$ measurements. But in the interest
of simplifying the notation, we will continue to refer the signal
as $\kappa$-sparse and the number of required measurements as $2\kappa$.

\section{Proposed Scheme for Reduced Training Overhead\label{sec:Proposed-Scheme}}

We see from \prettyref{fig:ChannelPlot}c that the time domain channel
$\textbf{h}$ only has a few significant taps, making $\textbf{h}$
a sparse vector. The measurement matrix can then be chosen as the
Fourier matrix $\textbf{F}$, resulting in the frequency domain channel
measurements: $\textbf{H}=\textbf{F}\textbf{h}$. We could then apply
compressed sensing theory and measure only a few elements of $\textbf{H}$,
from which we can recover $\textbf{h}$ using CoSaMP. Denoting the
sparsity of $\textbf{h}$ by $\kappa$, we use a sensing (or sampling
or selection) matrix $\textbf{S}$ that is $2\kappa\times N_{DFT}$
whose rows are unit vectors (i.e., the matrix has entries that are
either $1$ or $0$ ) to pick rows of $\textbf{F}$. The $2\kappa\times1$
measured channel vector is then $\textbf{H}_{2\kappa}=\textbf{S\textbf{F\textbf{h}}}$.
\begin{figure}[tbh]
\begin{centering}
\includegraphics{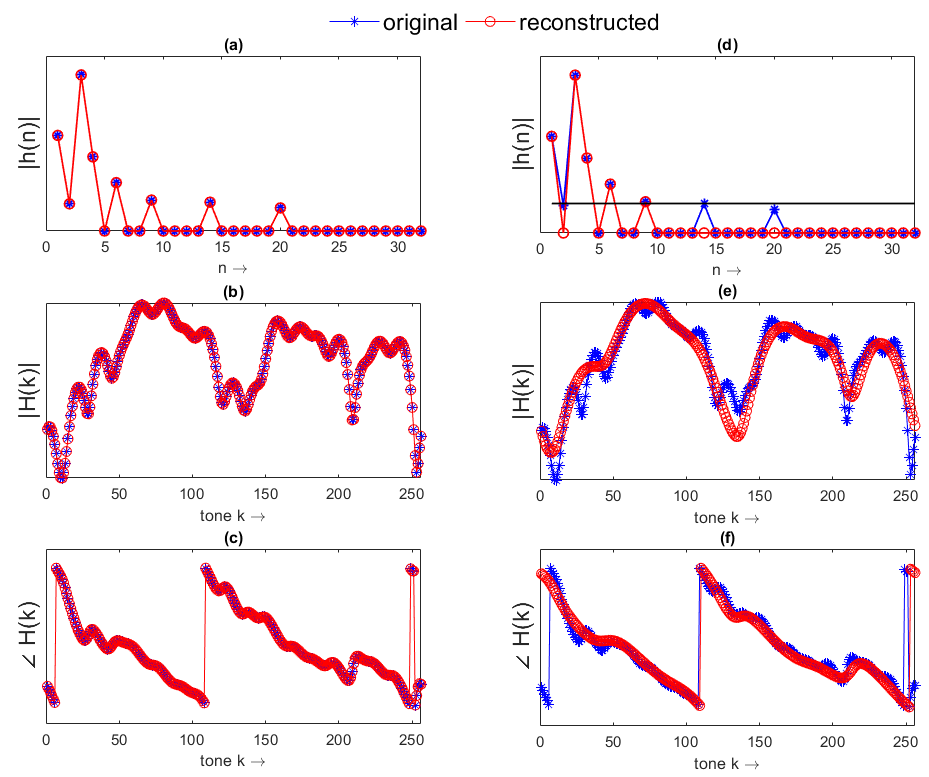}
\par\end{centering}
\caption{Recovery of $\textbf{h}$ using Compressed Sensing\label{fig:1DCompressedSensing}}
\end{figure}

\prettyref{fig:1DCompressedSensing} sub-plots (a--c) shows the perfect
recovery that we can achieve with CoSaMP using just $32$ random elements
of the frequency domain channel vector $\textbf{H}$ instead of the
full $256$. \prettyref{fig:1DCompressedSensing} sub-plots (d--f)
show the recovery if we zero out the elements of $\textbf{h}$ that
are below a threshold, and since this reduces the number of non-zero
entries in $\textbf{h}$, we only needed $24$ measurements out of
the $256$ length $\textbf{H}$ vector for the reconstruction.

One way of extending this to MIMO channel is to stack the channel
impulse responses for every TX-RX path. Denoting the $N_{DFT}\times1$
time domain channel vector between transmit antenna $\ell$ and receive
antenna $m$ to be $\textbf{h}_{\ell m}$, we will have the $N_{t}N_{r}N_{DFT}\times1$
stacked channel vector $\check{\textbf{h}}=[\textbf{h}_{11}^{T}\textbf{h}_{21}^{T}\cdots\textbf{h}_{\ell1}^{T}\ldots\textbf{h}_{N_{t}N_{r}}^{T}]^{T}.$
With $N=N_{t}N_{r}N_{DFT}$ we could use a $N\times N$ Fourier matrix
$\mathbf{F}_{N}$ to compute the frequency domain measurement vector
$\mathbf{H}_{1D}=\mathbf{F}_{N}\check{\mathbf{h}},$but we note here
that if $\textbf{h}_{\ell m}$ is $\kappa$-sparse, then the sparsity
of $\check{\textbf{h}}$ is $N_{t}N_{r}\kappa$. The number of measurements
required therefore scales linearly with the number of spatial dimensions
$N_{s}=N_{t}N_{r}$. From here on, we construct the $N_{DFT}\times N_{s}$
time domain channel matrix $[h(n,s)]=[\begin{array}{cccccc}
\textbf{h}_{11} & \textbf{h}_{21} & \cdots & \textbf{h}_{\ell m} & \cdots & \textbf{h}_{N_{t}N_{r}}\end{array}]$ where the columns $\mathbf{h}_{\ell m}$ are the channel impulse
responses for each TX-RX path.
\begin{figure}[tbh]
\begin{centering}
\includegraphics{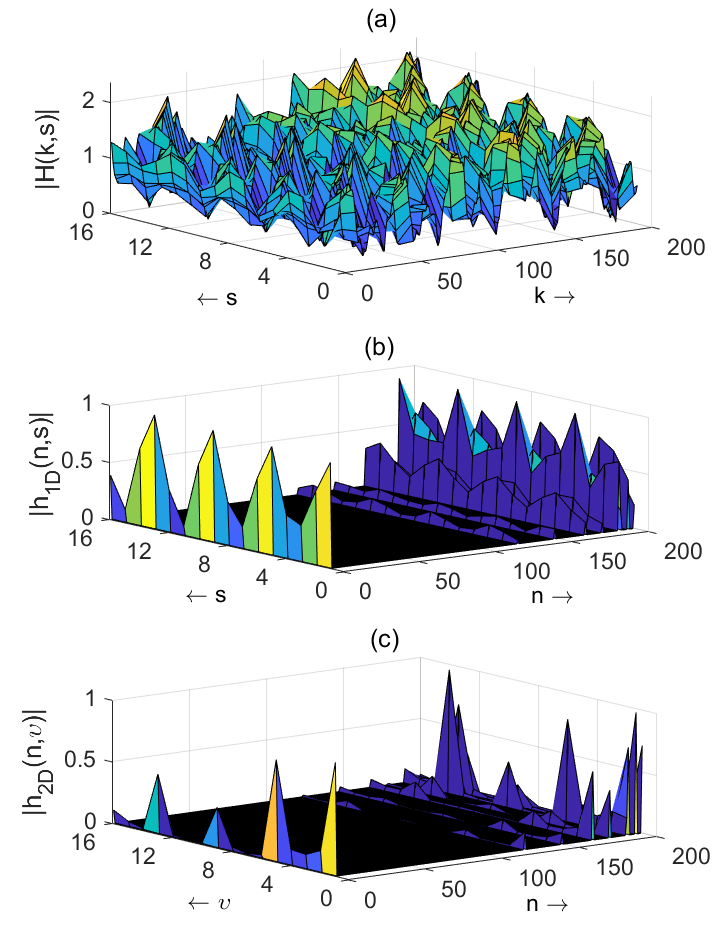}
\par\end{centering}
\caption{Channel Transformations \label{fig:2D-Chan}}
\end{figure}

We see from \prettyref{fig:2D-Chan} that the dense frequency domain
channel $[H(k,s)]$ shown in (a) transforms to $[h(n,s)]=\mathbf{F}_{N_{DFT}}^{H}[H(k,s)]$
shown in (b) and we see that there are still quite a few non-zero
elements in $[h(n,s)]$. This is because the compression is achieved
across tones, but there is no compression across the spatial dimension.
For large number of antennas, the correlation between antennas are
low if they are separated by a large distance ($\gg\lambda/2,$where
$f_{c}=c/\lambda$ is the carrier frequency), but for $f_{c}=2.4$
GHz, $\lambda=12.5$cm. If we have to place 8 or 16 antennas, it will
be quite difficult to achieve significant antenna spacing in an AP,
thus resulting in correlated antennas. We could exploit this correlation
to achieve compression in spatial dimension as well. To this end,
we compute the 2D inverse Fourier Transform in \prettyref{eq:2D-IFFT},
and see that the result has more sparsity as seen in \prettyref{fig:2D-Chan}c.
If we rearrange the matrix $[h(n,\upsilon)]$ to form a vector $\mathring{\mathbf{h}}=[\begin{array}{cccc}
\mathbf{h}_{1} & \mathbf{h}_{2} & \cdots & \mathbf{h}_{N_{DFT}}\end{array}]^{T}$where $\mathbf{h}_{n}$ denotes the $n$\textsuperscript{th}row of
$[h(n,\upsilon)]$, then we can rewrite \prettyref{eq:2D-FFT} to
get \prettyref{eq:kronEqn}. Here $\otimes$ is the Kronecker product
of matrices $\mathbf{F}_{N_{DFT}}$ and $\mathbf{F}_{N_{s}}$. Equation
\prettyref{eq:kronEqn} immediately gives the formulation we required
to apply Compressed Sensing, with $\mathring{\mathbf{h}}$ being the
$\kappa$-sparse vector, $\left(\mathbf{F}_{N_{DFT}}\otimes\mathbf{F}_{N_{s}}\right)$
being the $N_{DFT}N_{s}\times N_{DFT}N_{s}$ measurement matrix, and
$\mathring{\mathbf{H}}$ is the measured signal vector. To recover
$\mathring{\mathbf{h}}$, we only need $2\kappa$ measurements of
$\mathring{\mathbf{H}}$.
\begin{eqnarray}
[h(n,\upsilon)] & = & \mathbf{F}_{N_{DFT}}^{H}[H(k,s)]\mathbf{F}_{N_{s}}^{H}\label{eq:2D-IFFT}\\{}
[H(k,s)] & = & \mathbf{F}_{N_{DFT}}[h(n,\upsilon)]\mathbf{F}_{N_{s}}\label{eq:2D-FFT}\\
\mathring{\mathbf{H}} & = & \left(\mathbf{F}_{N_{DFT}}\otimes\mathbf{F}_{N_{s}}\right)\cdot\mathring{\mathbf{h}}\label{eq:kronEqn}
\end{eqnarray}

\begin{figure}[tbh]
\begin{centering}
\includegraphics{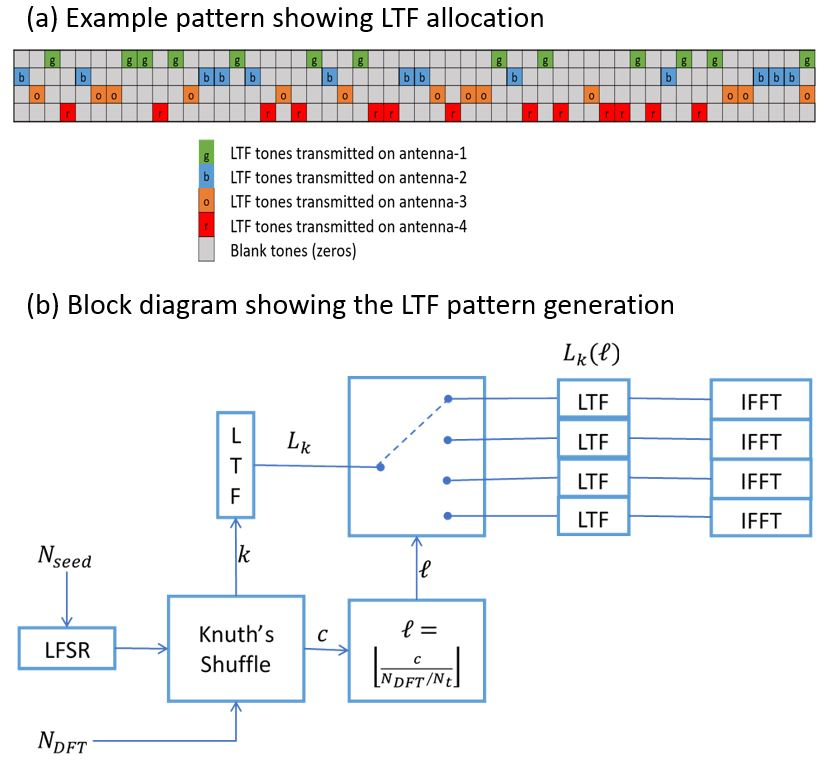}
\par\end{centering}
\caption{LTF allocation and pattern generation\label{fig:LTF-allocation}}
\end{figure}

Since we only need $2\kappa$ measurements of the channel $[H(k,s)]$,
it is possible to puncture the LTFs at the transmitter and only transmit
it for a few random tones, but note that we need at least one whole
symbol to do FFT at the receiver. We propose to remove the $\mathbf{P}$
matrix and transmit the LTF symbol $L_{k}$ for a given tone $k$
only from one of the $N_{t}$ antennas, with other antennas transmitting
zeros. \prettyref{fig:LTF-allocation}a shows an example pattern for
$52$ tones and $N_{t}=4$ transmit antennas, with $13$ LTF tones
transmitted per antenna. The LTF tone locations for each antenna is
non-overlapping with the LTF tone location for the other antennas.
These random LTF locations for each antenna can be arrived at by using
Knuth's shuffling algorithm \citep{Knuth-Shuffle} together with an
LFSR to generate a random permutation of tone indices $k=1,\cdots,N_{DFT}$,
which we then partition equally across all transmit antennas as shown
in \prettyref{fig:LTF-allocation}b. The seed for the LFSR can be
communicated by the beamformer to the beamformee as part of the association
handshake process or in NDPA.

\begin{figure}[tbh]
\begin{centering}
\includegraphics{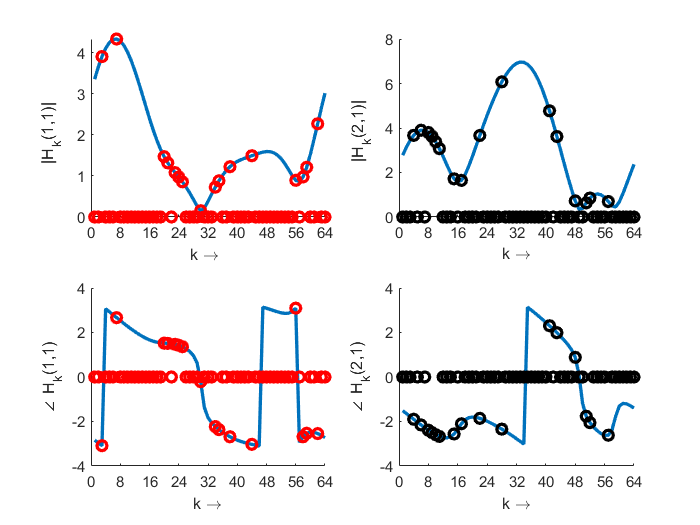}
\par\end{centering}
\caption{Channel estimation from punctured LTF\label{fig:PuncturedLTFChanEst}}

\end{figure}

Note from \prettyref{fig:LTF-allocation}b that the puncturing of
the LTFs happen in the frequency domain (before IFFT) at the transmitter.
\prettyref{fig:PuncturedLTFChanEst} shows the resulting channel estimation
for a $2\times1$ system where the LTF symbols for each transmit antenna
has been allocated using the proposed method. We see that we can only
estimate the channel for those tones in which LTF has been transmitted,
and the estimated channel (plotted as circles in \prettyref{fig:LTF-allocation})
matches the original channel (plotted with solid lines) for those
tones. 

Another added advantage of removing $\mathbf{P}$ matrix is that the
LTF on any given antenna can be transmitted with higher power since
all of the total power is allocated to one antenna and not divided
across the $N_{t}$ antennas as it would have been for LTF transmission
with the $\mathbf{P}$ matrix. This will also result in better SNR
for channel estimation at the receiver.

At the receiver, we calculate the LTF locations based on the LFSR
seed, estimate the channel $H_{k}(\ell,m)$ on those locations and
send them back in feedback. If the number of measurements $N_{\kappa}$
required is less than $N_{DFT}$, we will use only one LTF symbol
for all $N_{t}$ antennas. If $N_{\kappa}>N_{DFT}$, we need $\lceil N_{\kappa}/N_{DFT}\rceil$
LTF symbols. Note that $l_{1}$ minimization requires at least $2\kappa$
measurements to recover a $\kappa$-sparse real vector, and for complex
vector we have $N_{\kappa}\geq4\kappa$. We can reduce the number
of feedback bits if we just send the $\kappa$ non-zero complex time-domain
channel taps and its locations, but we not only need significant processing
power at the receiver to estimate the time domain channel by computing
$N_{t}N_{r}$ FFTs, we also need access to the channel matrix for
all the tones. Instead, this method allows us to reduce the number
of LTFs transmitted while at the same time avoiding computationally
intensive processing at the receiver. Usually, the beamformer is an
access point that is plugged into a wall socket, and the beamformee
is a battery operated device and it is beneficial to reduce computations
at the beamformee to save power consumption. The beamformer can then
run an $l_{1}$minimization algorithm such as CoSaMP on the received
feedback to recover the full channel vector, which can be used to
then compute the SVD precoder. 
\begin{figure}[tbh]
\begin{centering}
\includegraphics{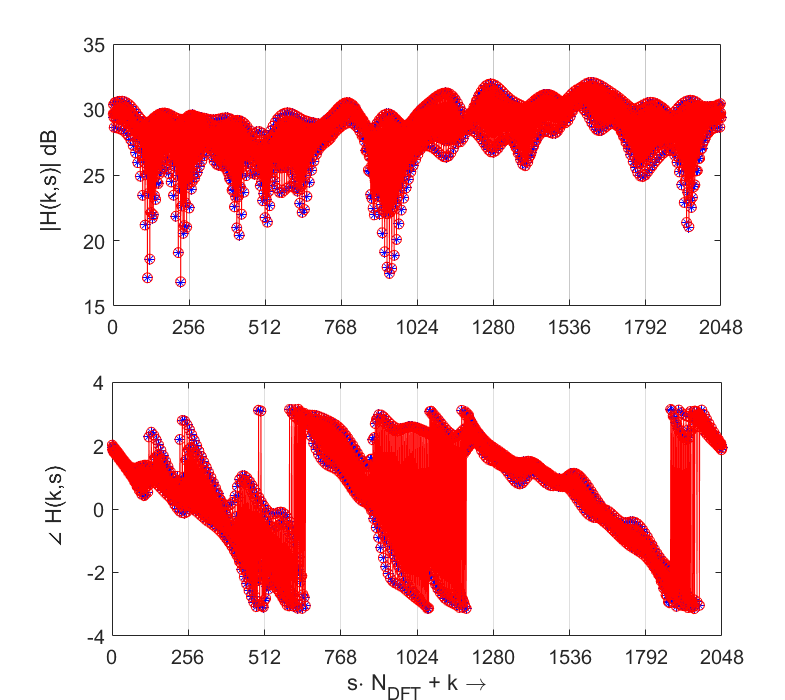}
\par\end{centering}
\caption{CoSaMP recovery for $N_{DFT}=256$, $N_{t}=4$, $N_{r}=2$, $N_{\kappa}=200$
with the proposed scheme.\label{fig:CoSaMP-Result}}
\end{figure}

\prettyref{fig:CoSaMP-Result}a shows an example of recovery using
\prettyref{eq:kronEqn}. We used $N_{DFT}=256$, $N_{t}=4$, $N_{r}=2$
and the channel vector $\mathring{\mathbf{h}}$ in \prettyref{eq:kronEqn}
has sparsity $\kappa\leq50$, the number of measurements we need is
then $N_{\kappa}=4\kappa=200$. The blue circles are the original
channel vector and the red circles are the recovered channel vector
from $N_{\kappa}$ measurements.

\begin{figure}[tbh]
\begin{centering}
\includegraphics{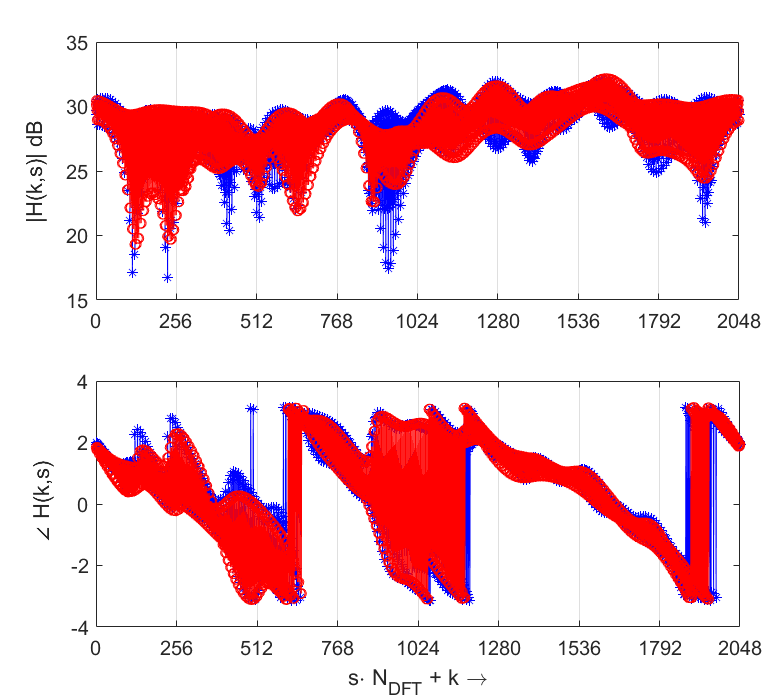}
\par\end{centering}
\caption{CoSaMP recovery for $N_{DFT}=256$, $N_{t}=4$, $N_{r}=2$, $N_{\kappa}=160$
with thresholding to reduce $\kappa\protect\leq35$.\label{fig:CoSaMP-result-thresholding}}

\end{figure}
\begin{figure}[tbh]
\begin{centering}
\includegraphics{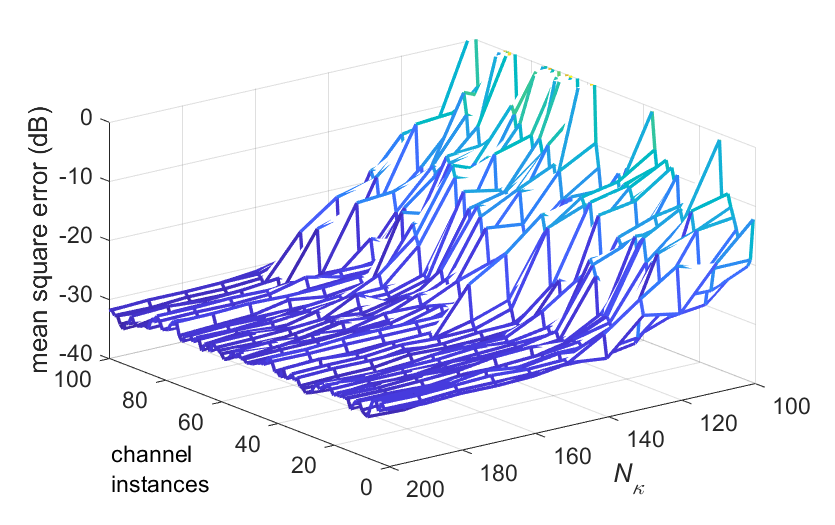}
\par\end{centering}
\caption{Mean Square Error for different values of $N_{\kappa}$\label{fig:Mean-Square-Error}}
\end{figure}

We can reduce the sparsity even further by ignoring all channel taps
in $\mathring{\mathbf{h}}$ that are $30$ dB below the peak, and
this allows us to reduce the number of measurements to $N_{\kappa}=160$
since the sparsity reduces to $\kappa<35$. There is however a resulting
error in the recovered channel as seen in \prettyref{fig:CoSaMP-Result},
since we have ignored some channel taps, but as long as the mean square
error is acceptable, this method provides us a way to trade-off the
feedback accuracy with number of feedback bits. The mean squared error
between the original and reconstructed channel vector for $\kappa=35$
is plotted in \prettyref{fig:Mean-Square-Error} for different channel
vectors and $N_{\kappa}.$ Note that ignoring the channel taps below
a threshold is beneficial when the receiver is in low RSSI region
and the smaller taps will be dominated by noise due to low SNR.

\subsection{Implementation Complexity\label{sec:Implementation Complexity}}

At the heart of CoSaMP is the pseudo-inverse computation that is required
to solve the least square problem: $\mathbf{b}=\mathbf{\Phi}_{(T)}^{\dagger}\mathbf{y}$
(line \ref{CoSaMP-pseudo-inv} in Algorithm \prettyref{alg:CoSaMP}).
We could instead solve $\mathbf{\Phi}_{(T)}\mathbf{b}=\mathbf{y}$
directly by computing $\mathbf{\Phi}_{(T)}^{H}\mathbf{\Phi}_{(T)}\mathbf{b}=\mathbf{\Phi}_{(T)}^{H}\mathbf{y}$
and finding the Cholesky decomposition $\mathbf{\Phi}_{(T)}^{H}\mathbf{\Phi}_{(T)}=\mathbf{L}\mathbf{L}^{H}$,
resulting in $\mathbf{L}\mathbf{L}^{H}\mathbf{b}=\mathbf{L}\mathbf{b}'=\mathbf{y}'$.
We can then first solve the triangular system of equations $\mathbf{L}\mathbf{b}'=\mathbf{y}'$
for $\mathbf{b}'$, and then solve the second triangular system of
equations $\mathbf{b}'=\mathbf{L}^{H}\mathbf{b}$ to get $\mathbf{b}$.
Efficient implementation of Cholesky decomposition has been well studied
in the literature and hardware implementation to exploit parallelization
have been explored \citep{Compressed-Sensing-Cholesky,OMP-Hardware-Impl,Cholesky-FPGA,Pseudo-Inverse-Updating}.
Cholesky decomposition is $\mathcal{O}(n^{3}/3)$ in complexity, and
for us $n\leq2\kappa$ since we pick at most $2\kappa$ columns out
of $N$ in the $N_{\kappa}\times N$ measurement matrix $\Phi$. We
can do this because $\hat{\mathbf{x}}=\mathbf{\Phi}^{\dagger}\mathbf{y}$
is $\kappa$-sparse and we can remove the rows of $\mathbf{\Phi}^{\dagger}$
that correspond to zero elements of $\hat{\mathbf{x}}$. For a $N_{DFT}=256$,
$N_{t}=4$, $N_{r}=2$ system with a sparsity factor of $\kappa=50$,
even though the channel vector length is $256\times4\times2=2048$,
the complexity of the least squares step in CoSaMP depends only on
the sparsity factor $\kappa$. Note that we can control the sparsity
factor $\kappa$ by choosing a threshold and ignoring all channel
taps below this threshold and doing so allows us the trade-off between
the complexity and the mean square error.

Main factors that contribute to CoSaMP complexity are lines 4, 7 and
10 in Algorithm \prettyref{alg:CoSaMP}. We have a matrix vector multiplication
$(N\times N_{\kappa})\cdot(N_{\kappa}\times1)$ in line 4 that requires
$NN_{\kappa}$ complex MAC operations. The pseudo-inverse in line
7 can be broken down into computing $\mathbf{\Phi}_{(T)}^{H}\mathbf{\Phi}_{(T)}$
which is a $(2\kappa\times N_{\kappa})\cdot(N_{\kappa}\times2\kappa)$
matrix multiplication requiring $N_{\kappa}(2\kappa)^{2}$ complex
MAC, and Cholesky decomposition requiring about $(2\kappa)^{3}$ operations.
Line 10 is another matrix vector multiplication $(N_{\kappa}\times2\kappa)\cdot(2\kappa\times1)$
with complexity $N_{\kappa}(2\kappa).$ Summing up, we end up with
a complexity of $\mathcal{O}(NN_{\kappa}+N_{\kappa}(2\kappa)+N_{\kappa}(2\kappa)^{2}+(2\kappa)^{3})$
which plugging in the numbers $N=2048$, $N_{\kappa}=256$ and $\kappa=50$
gives us about $4\cdot10^{6}$ complex MAC operations. In a $2$ GHz
computer, that translates to $2$ milliseconds for one CoSaMP iteration.
We expect about 10x speed up (see \citep{Cholesky-FPGA}) if dedicated
hardware is designed for CoSaMP exploiting parallel architectures,
and that gives us around $200\mu s$. This approximate calculation
aligns with the results in \citep{OMP-Hardware-Impl} where the authors
have measured a run time of $340\mu s$ on a 120 MHz Xilinx FPGA.

\section{Summary\label{sec:Summary}}

In this paper, we have outlined a proposal to reduce the training
overhead for MIMO channel estimation and feedback by using compressed
sensing framework for subsampling channel measurement and recovering
the full channel from the reduced set of channel estimates. To achieve
this, we presented a novel way of transforming the channel across
both frequency and spatial dimensions, using a 2D FFT which we then
reformulate to fit into the compressed sensing model. We have also
described a scheme by which the LTF locations for each antenna can
be generated. The simulation results presented here show that this
scheme works as intended and achieves the expected results. Complexity
analysis shows that with hardware accelerators to implement the key
steps in CoSaMP, we should achieve reconstruction in less than $0.5$
millisecond. For WLAN however, taking $0.5ms$ per user will result
in significant latency between the channel measurement time and the
time the feedback is used for precoding. This might still be sufficient
for slow changing channels, but we still require advances in hardware
architecture to be able to achieve the required reconstruction time
of less than $100\mu s$. But nevertheless, this study outlines a
viable method for reducing training and feedback overhead in WLAN
and can be used to build upon as Compressed Sensing becomes a more
mainstream topic in Wireless Systems.


\begin{thebibliography}{99}
\bibitem{OFDM-Reference}J. Terry, J. Heiskala, \textit{OFDM Wireless
LANs: A Theoretical and Practical Guide}, Sams Publishing, USA, 2002.

\bibitem{IEEE-802-11-Std}\textquotedbl IEEE Std. no. 802.11-2016\textquotedbl ,
\textquotedbl Standard for information technology -- specific requirements
-- part 11: Wireless LAN medium access control (MAC) and physical
layer (PHY) specifications\textquotedbl , Dec. 2016.

\bibitem{Compressed-Sensing} D. L. Donoho, \textquotedbl\textit{Compressed
sensing},\textquotedbl{} in IEEE Transactions on Information Theory,
vol. 52, no. 4, pp. 1289-1306, April 2006.

\bibitem{Candes-Intro-Compressive-Sampling} E. J. Candes, M. B. Wakin,
\textquotedbl\textit{An Introduction To Compressive Sampling},\textquotedbl{}
in IEEE Signal Processing Magazine, vol. 25, no. 2, pp. 21-30, March
2008.

\bibitem{TGnChannelModel}V. Erceg, L. Shumacher, P. Kyritsi, et al.,
\textquotedblleft TGn Channel Models,\textquotedblright{} IEEE 802.11-03/940r4,
10 May 2004.

\bibitem{CoSaMP}D. Needell, J. A. Tropp, \textquotedbl\textit{Cosamp:
Iterative signal recovery from incomplete and inaccurate samples},\textquotedbl{}
Appl. Comp. Harmonic Anal., 2008.

\bibitem{Knuth-Shuffle}Knuth, Donald E. (1969). \textit{Seminumerical
algorithms. The Art of Computer Programming. 2}. Reading, MA: Addison--Wesley.
pp. 139--140.

\bibitem{Compressed-Sensing-Cholesky}D. Yang, G. D. Peterson, H.
Li, \textquotedblleft \textit{Compressed sensing and Cholesky decomposition
on FPGAs and GPUs},\textquotedblright{} Parallel Computing. 2012.

\bibitem{OMP-Hardware-Impl}H. Rabah, A. Amira, B. K. Mohanty, S.
Almaadeed, P. K. Meher, \textquotedblleft \textit{FPGA Implementation
of Orthogonal Matching Pursuit for Compressive Sensing Reconstruction},\textquotedblright{}
in IEEE Transactions on Very Large Scale Integration (VLSI) Systems,
Oct. 2015.

\bibitem{Cholesky-FPGA}D. Yang, G. D. Peterson, H. Li, \textquotedblleft \textit{High
performance reconfigurable computing for Cholesky decomposition},\textquotedblright{}
in Symposium on Application Accelerators in High Performance Computing
(SAAHPC), Jul. 2009.

\bibitem{Pseudo-Inverse-Updating}C. T. Pan, R. J. Plemmons, ``\textit{Least
squares modifications with inverse factorizations: parallel implications},''
Journal of Computational and Applied Mathematics 27, 1989.
\end{thebibliography}
\end{document}